\documentstyle[psfig]{europhys}

\def\And{{\rm and\ }}

\newif\ifboo \boofalse

\def\Review#1{\boofalse{\it #1},}

\def\Vol#1{\ifboo Vol. {\bf #1}\else{\bf #1}\fi}
\def\Year#1{\ifboo #1\else(#1)\fi}

\def\Page#1{\ifboo {\rm p. #1}\else{\rm #1}\fi}

\newcommand{\lesssim}{
\,\raisebox{0.35ex}{$<$}
\hspace{-1.7ex}\raisebox{-0.65ex}{$\sim$}\,
}
\input{tcilatex}

\begin{document}

\euro{XX}{X}{1-$\infty$}{2002} \Date{1 February 2002}
\shorttitle{D. A. GARANIN \And R. SCHILLING: INVERSE PROBLEM FOR
THE LANDAU-ZENER EFFECT}

\bibliographystyle{prsty}

\title{Inverse problem for the Landau-Zener effect}
\author{D. A. Garanin \And R. Schilling}
\institute{Institut f\"ur Physik, Universit\"at Mainz, D-55099
Mainz, Germany}
\pacs{ \Pacs{03}{65.-w}{Quantum mechanics}
 \Pacs{75}{10.Jm}{Quantized spin models}
\Pacs{03}{67.Lx}{Quantum computation} }
\maketitle
\vspace{-1cm}
\begin{abstract}
We consider the inverse Landau-Zener problem which consists in
finding the energy-sweep functions $W(t)\equiv
\varepsilon_1(t)-\varepsilon_2(t)$ resulting in the required time
dependences of the level populations for a two-level system
crossing the resonance one or more times during the sweep.
We find
sweep functions of particular forms that let manipulate the system
in a required way, including complete switching from the state 1
to the state 2 and preparing the system at the exact ground and
excited states at resonance.
\end{abstract}

\vspace{-1cm}
 The Landau-Zener (LZ) effect
\cite{lan32,zen32,dobzve97}, which consists in the
quantum-mechanical transition between the two time-dependent
levels of a two-level system (states 1 and 2) caused by crossing
the resonance, is a well known phenomenon in many areas of
physics.
Well known applications of the LZ effect are those to molecular
collision and  dissociation\cite{lanlif3,crohug77}.
Recently, LZ effect has been used to measure the tunnel splitting
in molecular magnets whereby detecting topological interference
effects \cite{werses99} which were predicted theoretically
\cite{losdivigri92,delhen92,garg93}.
If the system was in state 1 before crossing the resonance, then,
for a constant sweep rate $v$, the probability to stay in this
state after crossing the resonance is given by
\begin{equation}\label{LZProbab}
P=\exp \left( -\frac{\pi \Delta ^{2}}{2\hbar v}\right),
\end{equation}
where $v\equiv  |\dot{W}(t)|$, $W(t)\equiv
\varepsilon_{1}(t)-\varepsilon _{2}(t)$, $\varepsilon
_{1,2}(t)=\langle \psi _{1,2}|\hat{H}|\psi _{1,2}\rangle$, $\Delta
\equiv 2\langle \psi _{1}|\hat{H}|\psi _{2}\rangle$, $W(t)$
satisfies $W(\pm \infty )=\pm \infty $ and $\Delta $ is the tunnel
level splitting.
For {\em fast} energy sweep rates $v$ the system spends too little
time in the vicinity of the resonance so that the tunneling matrix
element $\langle \psi _{1}|\hat{H}|\psi_{2}\rangle $ cannot bring
the system into the state 2 (say, onto the other side of a
potential barrier), and the system remains in the initial state 1.
For {\em slow} sweep rates the probability to remain in the
initial state becomes exponentially small, and the system travels
into the state 2 remaining on the lower {\em exact} energy term
$\varepsilon _{-}(t)$ of
\begin{equation}
\varepsilon _{\pm }(t)=\frac{1}{2}\left[ \varepsilon
_{2}(t)+\varepsilon _{1}(t)\pm \sqrt{W^{2}(t)+\Delta ^{2}}\right]
. \label{epspmDef}
\end{equation}
The corresponding {\em adiabatic} time dependence of the
probability to be in state 1 is given by
\begin{equation}
|c_{1}(t)|^{2}=|\langle \psi _{1}|\psi _{-}(t)\rangle |^{2}=\frac{1}{2}\left( 1-%
\frac{W(t)}{\sqrt{W^{2}(t)+\Delta ^{2}}}\right) .
\label{ProbSlow}
\end{equation}
In the case where $\hbar v\sim \Delta ^{2}$ and $v$ is constant,
the solution of the Schr\"{o}dinger equation for $|c_{1}(t)|^{2}$
 can be found analytically \cite{lan32,zen32,dobzve97} and it is oscillating at $t>0$
 slowly approaching $P$ given by Eq.\
(\ref{LZProbab}) at $t\rightarrow \infty$.

%
%
%
%

The known results above are valid for the linear energy sweep,
$W(t)=vt.$ One can consider, however, other sweep functions $W(t)$
that behave {\em nonlinearly} in the vicinity of the resonance,
$W\sim \Delta $.
If the tunneling resonance is not too narrow and its position is
well defined, nonlinear sweeps of a controlled form can be
realized experimentally.
Although no general analytical solution of the Schr\"{o}dinger
equation is known for the LZ problem with an arbitrary $W(t),$ one
can look for the functions $W(t)$ that result in a desired time
dependence of the probabillity $|c_{1}(t)|^{2}$, thus solving the
{\em inverse} LZ problem.
It is the main motivation of this Letter to study this inverse
problem which, for instance, will allow to determine $W(t)$ needed
to manipulate and to prepare a quantum system in a specific state.

One class of solutions of the time-dependent Schr\"odinger
equation are smooth, non-oscillating functions of time that
describe the complete switching of the system from the state 1 to
the state 2, e.g., Eq.\ (\ref {ProbSlow}).
Another class of solutions are those starting in $\psi _{1}$ at
$t=-\infty $ off resonance and ending in a given superposition of
the exact states $\psi_{\pm }$ at resonance [$W(t)=0$ for $t\geq
0]$.
The most interesting variants of the latter are the system in the
ground state $\psi _{-}$ at resonance and in the excited state
$\psi_{+}$ at resonance.
These possibilities to manipulate the system in a required way
could prove to be useful for the quantum computing (QC).
Another interesting possibility is to sweep at a high rate
(nonlinearly in time) far from the resonance thus making the whole
process of crossing the resonance much shorter.
Exploring these new types of solution of the LZ problem is the aim
of this Letter.

\begin{figure}[t]
\unitlength1cm
\begin{picture}(7,5)
\psfig{file=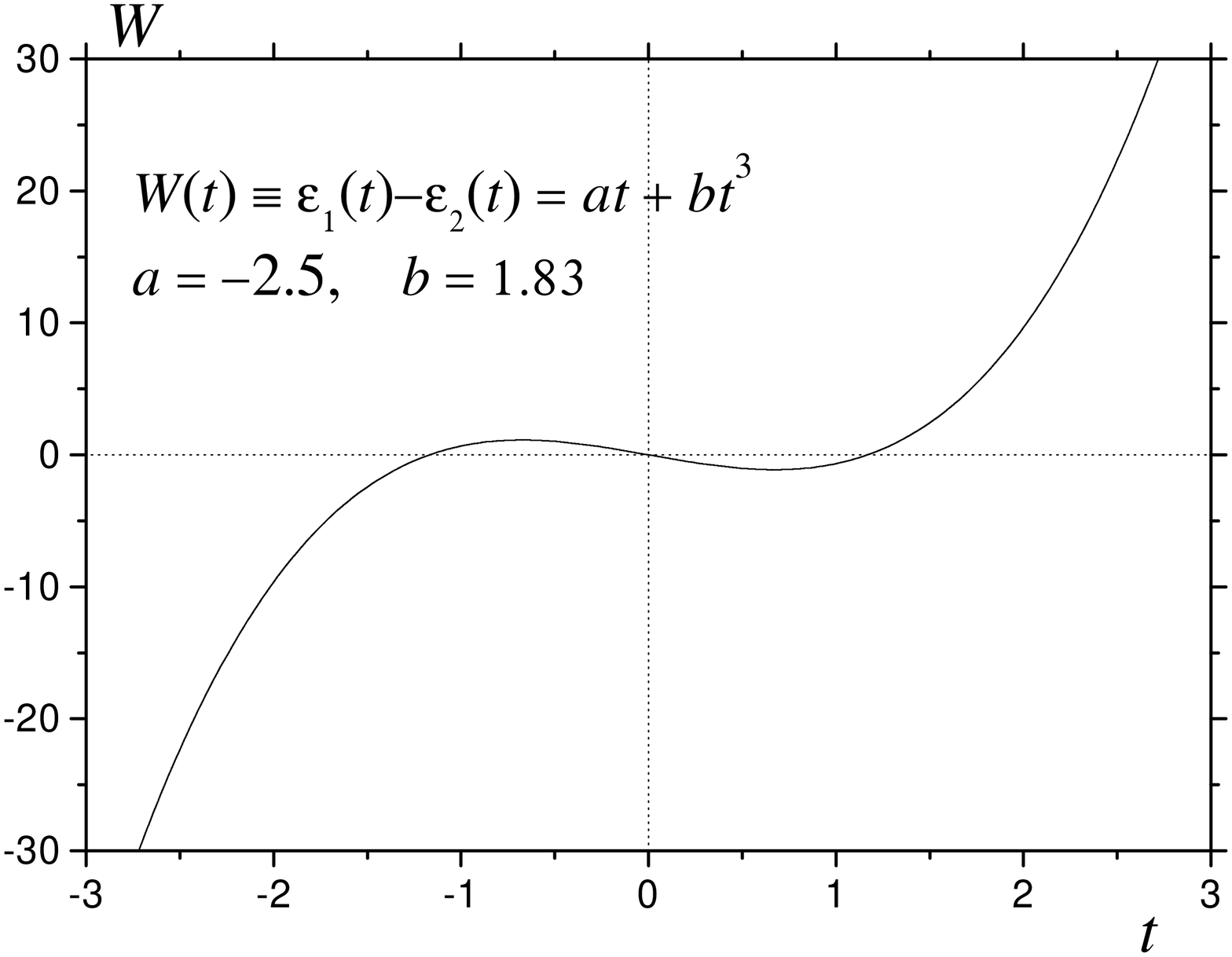,angle=-90,width=7.5cm}
\end{picture}
\begin{picture}(7,5)
\psfig{file=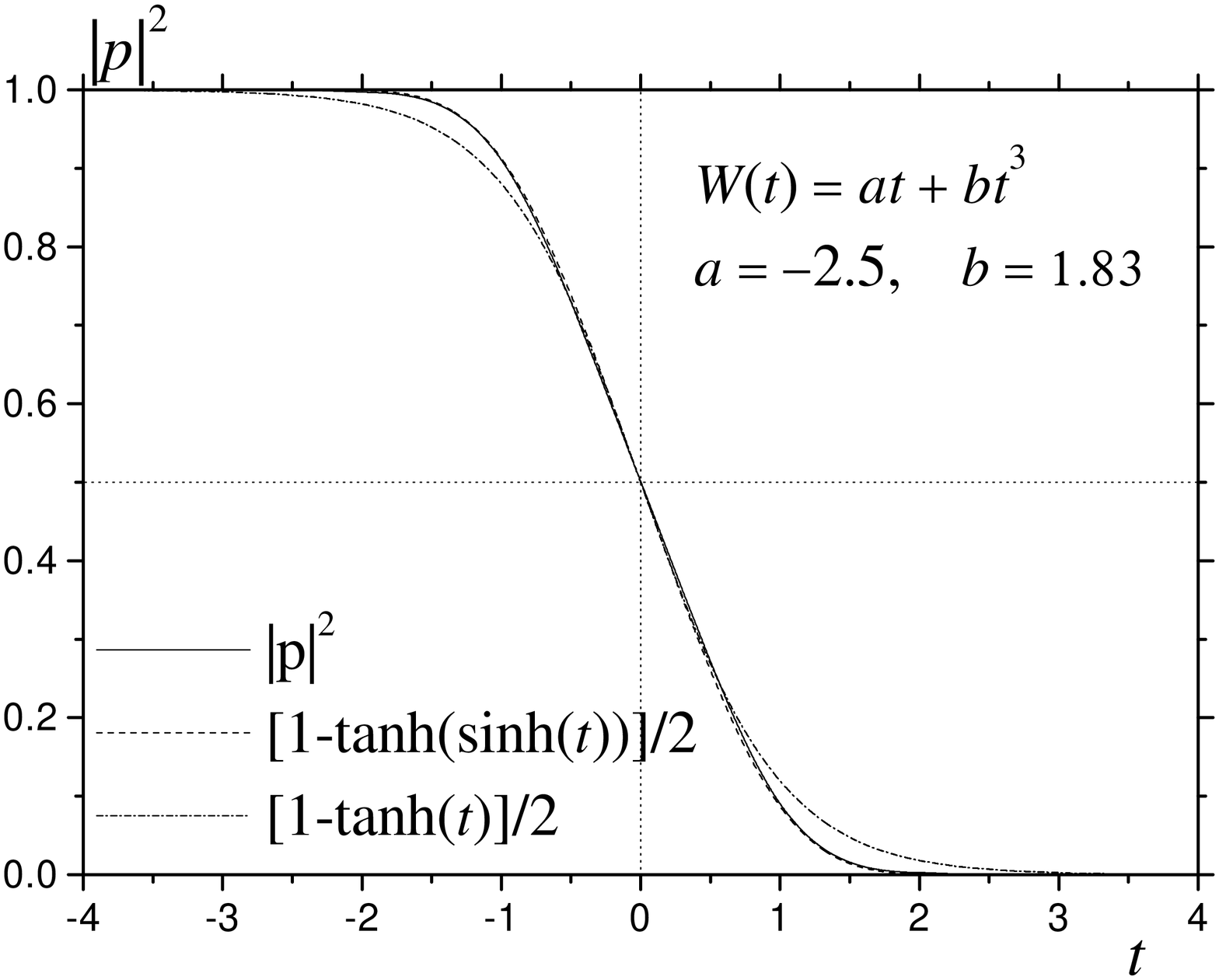,angle=-90,width=7.5cm}
\end{picture}
\caption{The cubic-parabola energy sweep with $W(t)=-2.5t+1.83t^3$
and $\Delta=\hbar=1$.
Time dependence of the probability $|c_1(t)|^2=|p(t)|^2$ can be
fitted with $|p|^2=[1-\tanh( \sinh t)]/2$ with graphic accuracy.
The function $|p|^2=[1-\tanh t]/2 $ is shown for comparison. }
\label{lzi-cub}
\end{figure}

To this end, we consider the Schr\"{o}dinger equation for the coefficients
in the wave function $\psi (t)=c_{1}(t)\psi _{1}+c_{2}(t)\psi _{2}$ that can
be written in the form
\begin{eqnarray}\label{pqEq}
\hbar \dot p(t) &=&-\frac{i}{2}\Delta q(t),\qquad \qquad \qquad
\quad
p(-\infty )=1  \nonumber \\
\hbar \dot q(t) &=& i W(t)q(t)-\frac{i}{2}\Delta p(t),\qquad
q(-\infty )=0,\qquad W(-\infty )=-\infty
\end{eqnarray}
with $p(t)\equiv e^{i\varepsilon _{1}t}c_{1}(t)$ and $q(t)\equiv
e^{i\varepsilon _{1}t}c_{2}(t)$.
We are looking for sweep functions $W(t)$ that result in a
required time dependence of the state populations.
 Before solving this inverse LZ problem
analytically, we will illustrate numerically the possibility of
time-symmetric solutions of Eq.\ (\ref{pqEq}) going from $
p(-\infty )=1$ to $p(\infty )=0$ for strongly nonlinear sweep
functions $ W(t)$.
 The simplest of these cases is the cubic-parabola sweep with $%
W(t)=-2.5t+1.83t^{3}$ where we have set $\Delta =\hbar =1$ in Eq.\ (\ref
{pqEq}).
The time dependence of the probability $|p(t)|^2$ shown in Fig.\
\ref{lzi-cub} can be fitted by $|p|^{2}(t)=[1-\tanh (\sinh
(t))]/2$ which is, however, not the analytical solution of Eq.\
(\ref{pqEq}).

To find analytical solutions of the inverse LZ problem, one should
rewrite Eq.\ (\ref{pqEq}) in terms of the probability $m\equiv
|p|^{2}$ and some phase variable.
We thus define
\begin{equation}
p=\sqrt{m}\exp (i\varphi _{p}),\qquad q=\sqrt{1-m}\exp (i\varphi _{q})
\label{pqviamphi}
\end{equation}
and substitute it in Eq.\ (\ref{pqEq}).
Evidently the global phase of the system
$(\varphi_{p}+\varphi_{q})/2$ is irrelevant and the resulting
Schr\"{o}dinger equation can be formulated in terms of $m$ and the
phase difference $\varphi =\varphi _{q}-\varphi _{p}$:
\begin{eqnarray}\label{mphiEq}
\hbar \dot m &=&\Delta \sqrt{m(1-m)}\sin \varphi   \nonumber \\
\hbar \dot\varphi &=&W+\frac{\Delta }{2}\frac{1-2m}{\sqrt{m(1-m)}}%
\cos \varphi .
\end{eqnarray}
This system of equations can be brought into a more elegant form by
introducing
\begin{equation}
x\equiv 2m-1,\qquad -1<x<1 , \label{xDef}
\end{equation}
\begin{equation}
y\equiv \arcsin x,\qquad -\pi /2<y<\pi /2,  \label{yDef}
\end{equation}
and the reduced variables
\begin{equation}
\tau \equiv \left( \Delta /\hbar \right) t,\qquad
\overline{W}\equiv W/\Delta   \label{tauDef}.
\end{equation}
Then Eqs.\ (\ref{mphiEq}) can be put into the reduced form
\begin{eqnarray}
\partial _{\tau }y &=&\sin \varphi   \nonumber \\
\partial _{\tau }\varphi  &=&\overline{W}-\tan y\cos \varphi .
\label{ypsiEqRed}
\end{eqnarray}
The boundary condition for this equation is $y(-\infty )=\pi /2$
whereas the value of $\varphi (-\infty )$ is irrelevant.
In general, far from the resonance $\varphi (\tau )$ strongly
oscillates with time and the derivative $\partial _{\tau }y$ is
not small.
Below we will consider special kinds of solutions of Eq.\
(\ref{ypsiEqRed}) which are characterized by a smooth,
non-oscillating dependence $\sin\varphi (\tau )$ and $\partial
_\tau{ }y(-\infty)=0$.
We will solve the inverse
problem for Eq.\ (\ref{ypsiEqRed}) which consists in finding $\overline{W}%
(\tau )$ resulting in a required function $x(\tau )$.
One easily obtains from Eq.\ (\ref{ypsiEqRed}) a general formula
\begin{equation}
\overline{W}(\tau )=\partial _{\tau }\varphi +\tan y\cos \varphi =\partial
_{\tau }^{2}y\left/ \cos \varphi \right. +\tan y\cos \varphi ,\qquad \cos
\varphi =\pm \sqrt{1-(\partial _{\tau }y)^{2}}.  \label{WbarRes}
\end{equation}
The sign in front of $\cos \varphi $ determines the sign of the
function $\overline{W}(\tau )$.
Since $|c_{1}(-\infty )|^{2}=1$ and thus $x(-\infty )=1$ and
$y(-\infty )=\pi /2$ one has $\tan y(-\infty )=\infty$.
To comply with the condition $W(-\infty )=-\infty $ one has to
choose, in general, the negative sign:
\begin{equation}
\overline{W}(\tau )=-\partial _{\tau }^{2}y\left/ \sqrt{1-(\partial _{\tau
}y)^{2}}\right. -\tan y\sqrt{1-(\partial _{\tau }y)^{2}}.  \label{WbarRes1}
\end{equation}

Before considering particular cases, we will comment on some
general properties of the functions $x(\tau )$ and
$\overline{W}(\tau )$.
 If $x(\tau )$ is an odd function of time
then $y(\tau )$ is also odd and $\partial _{\tau }y(\tau )$ is
even, whereas $\partial _{\tau }^{2}y(\tau )$ and $\tan y(\tau )$
are odd.
Then from the first of Eqs.\ (\ref{ypsiEqRed}) follows
that $\sin \varphi (\tau )$ is even.
If the sign of $\cos \varphi $ does not change, which is the
generic situation, $\cos \varphi (\tau )$ is even.
Then from Eq.\ (\ref{WbarRes}) follows that $\overline{W}(\tau )$
is odd.
In a similar way one can show that if $x(\tau )$ is even (the
system returns into the initial state $\psi _{1}$ after crossing
the resonance) then $\overline{W }(\tau )$ is even, too.

There is an important special case, however, which is realized if
the derivative $\partial _{\tau }y$ at some point $\tau _{0}$
attains its {\em maximal} value $|\partial _{\tau }y|=1$ which
follows from the first of Eqs.\ (\ref{ypsiEqRed}).
At this point $\cos \varphi $ turns to zero and changes its sign
to positive. The latter corresponds to the phase $ \varphi $ going
from the interval $\pi /2<\varphi <\pi $ into the interval $
0<\varphi <\pi /2$.
Note that $\partial _{\tau }^{2}y=0$ at $\tau =\tau _{0}$ thus
Eq.\ (\ref{WbarRes1}) does not diverge.
If $\tau _{0}=0$ then $\cos\varphi (\tau )$ is an odd function:
$\cos \varphi<0$ for $\tau<0$ and $\cos \varphi>0$ for $\tau>0$.
As a result, the parity of $\overline{W}(\tau)$ becomes inverse to
the parity of $x(\tau)$.
If $\tau _{0}\neq 0$ then for a symmetric function $x(\tau )$ the
similar happens at $\tau =-\tau_{0}$.
That is, $\cos \varphi (\tau)$ changes sign two times and is an
even function, thus the parity of $\overline{W}(\tau )$ coincides
with that of $x(\tau )$.
It should be noted that parity considerations for the direct LZ
problem do not apply: Symmetric forms of $\overline{W}(\tau )$ do
not necessarily result in symmetric solutions $x(\tau )$.
%
If, however, the solution $x(\tau )$ satisfies $x(\pm \infty)=\mp
1$ or $x(\pm \infty )=\pm 1$ and $\overline{W}(\tau)$ is time
symmetric, then $x(\tau)$ is also symmetric.
This follows from the time reversibility of the Schr\"{o}dinger
equation and the symmetry of the boundary values of $x(\tau )$.

Another implication of this analysis is the existence of the {\em
maximal} rate of the probability change.
We have seen that $|\partial _{\tau }y|\leq 1$, then from
$\partial _{\tau }x=\sqrt{1-x^{2}}\partial_{\tau }y$ follows
$|\partial _{\tau }x|\leq 1$.
This limitation is physically clear: Since tunneling between
$\psi_{1}$ and $\psi_{2}$ is caused by the matrix element $\Delta
\equiv 2\langle \psi_{1}|\hat{H}|\psi_{2}\rangle$, the latter sets
up the natural scale for the tunneling rate, see Eq.\
(\ref{tauDef}).
To completely tunnel from $\psi_{1}$ to $\psi_{2}$, the system
should spend some minimal time in the vicinity of the resonance,
that is, the sweep cannot be too fast.
Below we will discuss two most important families of the functions
$x(\tau )$ and corresponding $\overline{W}(\tau )$ which depend on
$\tau $ in the combination $u=a\tau$, the parameter $a$ being the
measure of the sweep rate.
We will see that there is a maximal value $a=a_{\max }$ for which
$|\partial _{\tau }y|=1$ at some $\tau =\tau _{0}$.
We will also see that although the functions $x(\tau )$ do
formally exist for $a>a_{\max }$, one cannot extend
$\overline{W}(\tau )$ into this region, as a result of the
physical limitation on the tunneling rate.
Thus solutions $x(\tau)$ with $a>a_{\max} $ are not realizable.

\begin{figure}[t]
\unitlength1cm
\begin{picture}(7,5)
\psfig{file=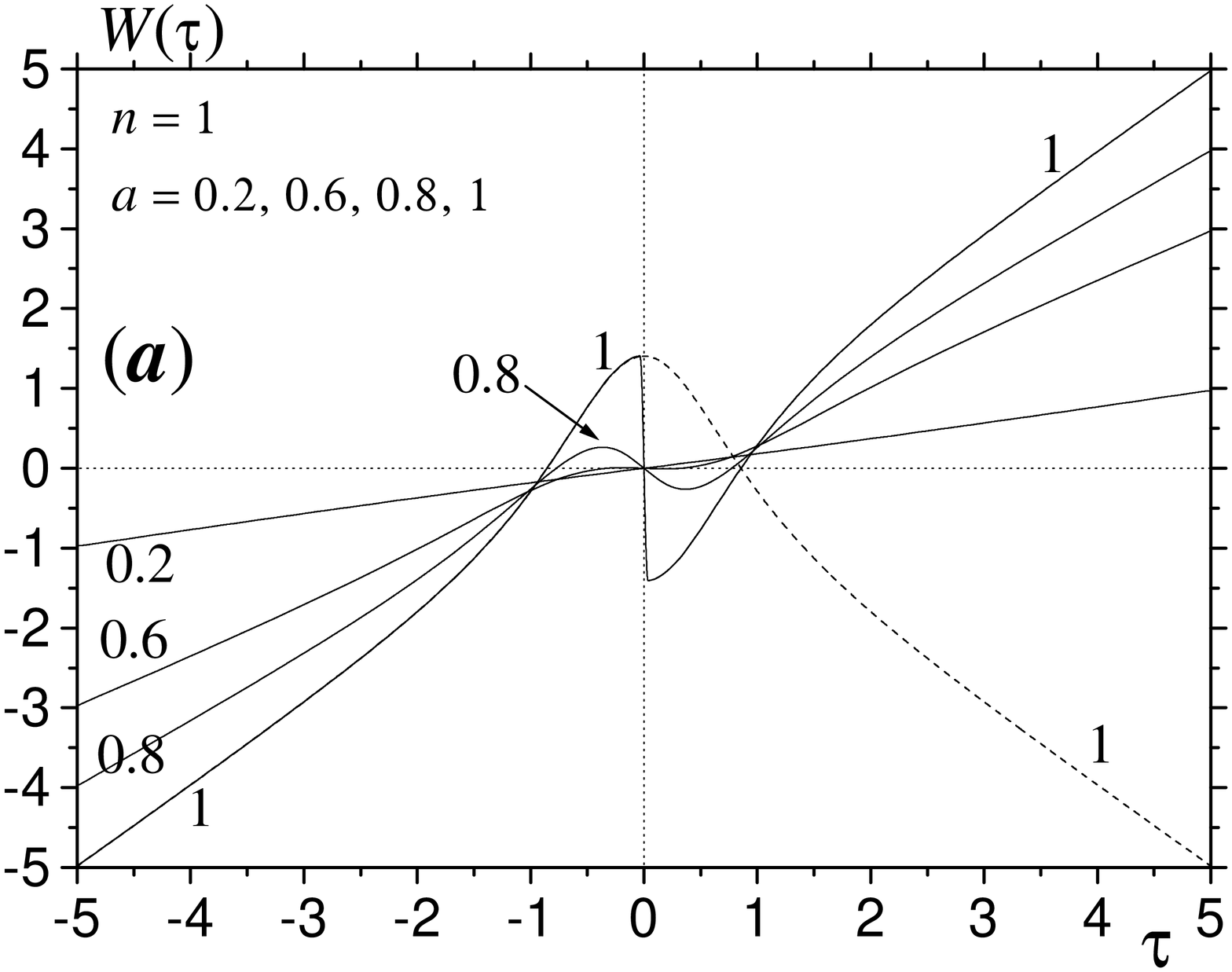,angle=-90,width=7.5cm}
\end{picture}
\begin{picture}(7,5)
\psfig{file=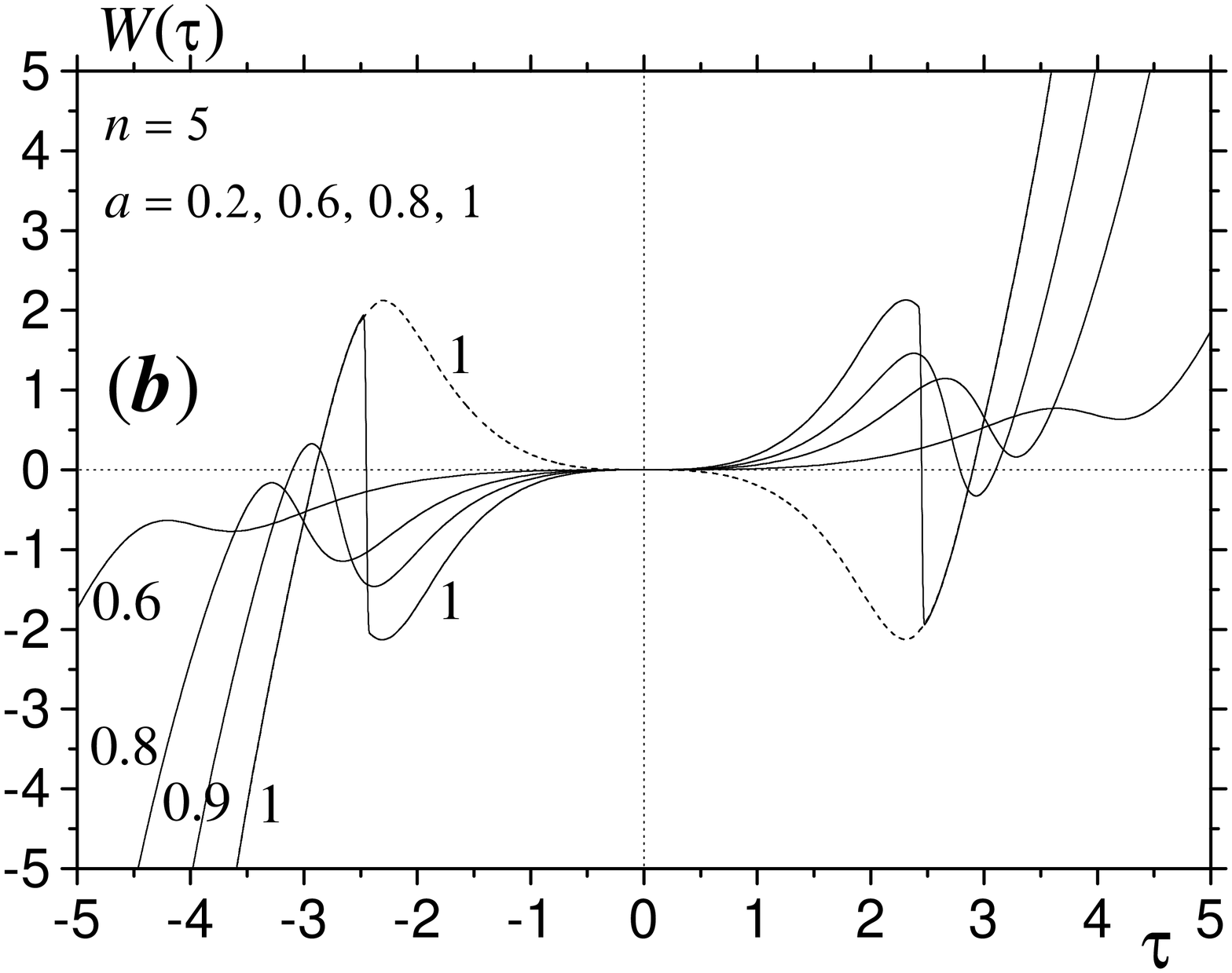,angle=-90,width=7.5cm}
\end{picture} \\
\begin{picture}(7,5)
\psfig{file=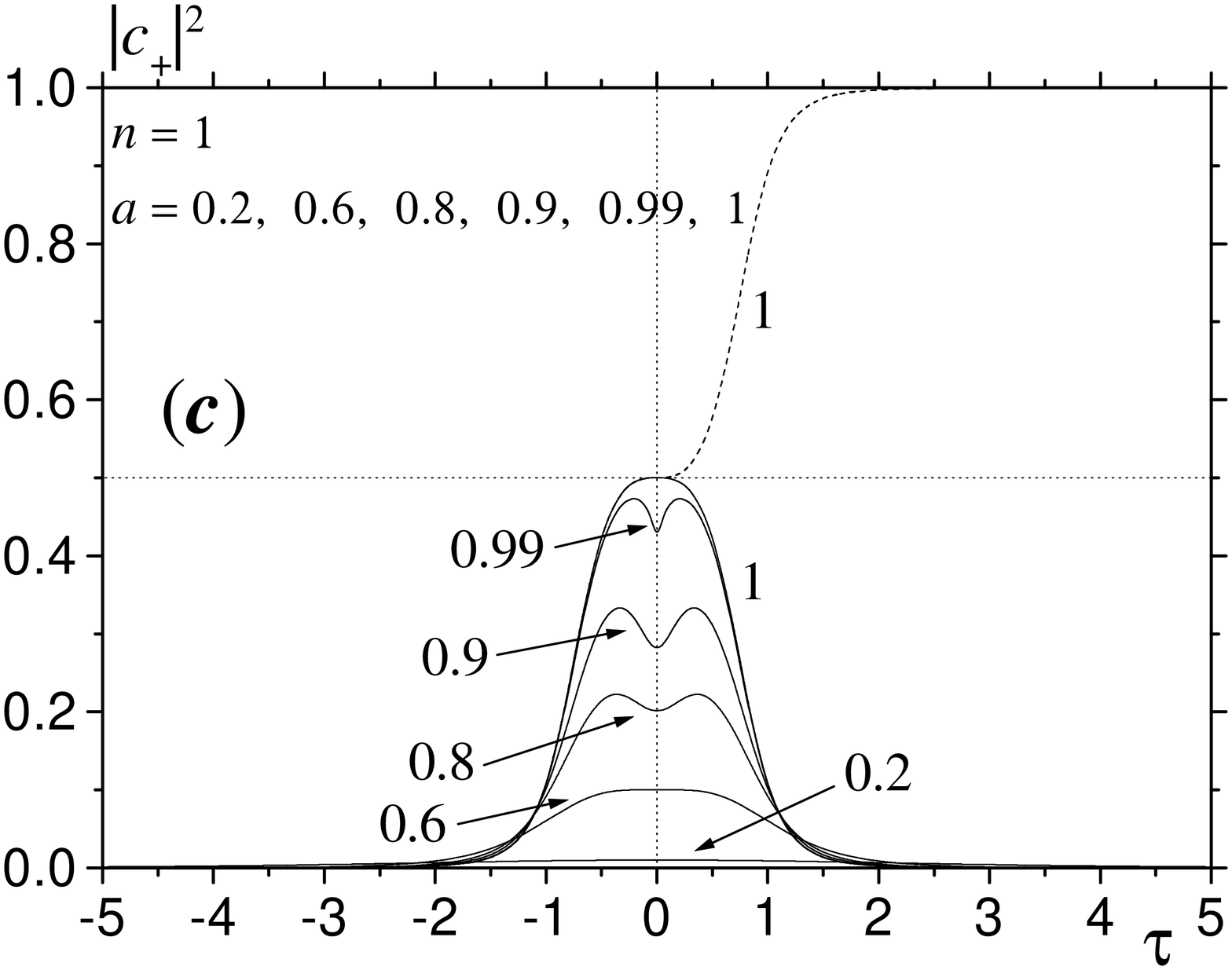,angle=-90,width=7.5cm}
%
\end{picture}
\begin{picture}(7,5)
\psfig{file=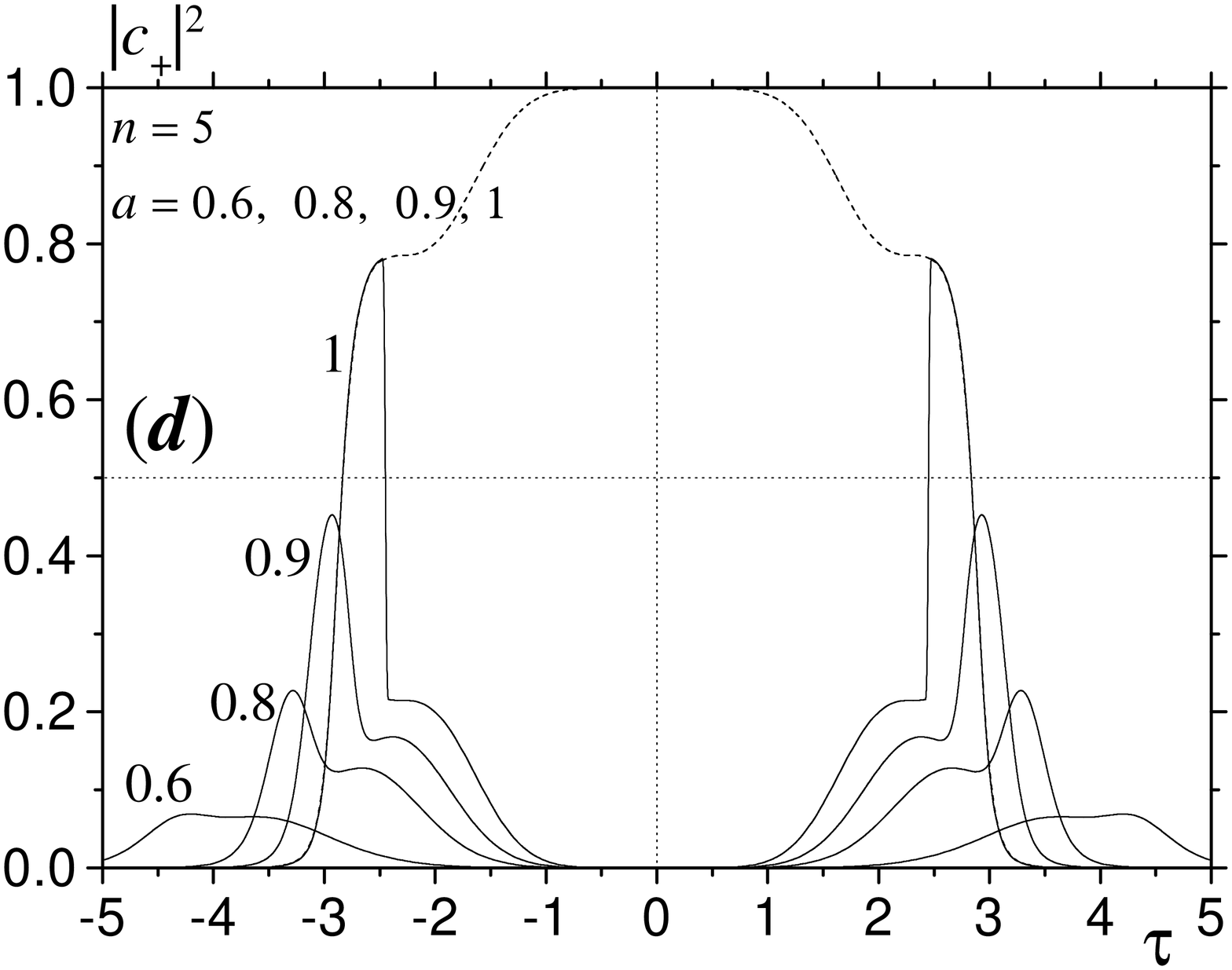,angle=-90,width=7.5cm}
\end{picture}
\caption{ \label{lzi-n1and5}
$a$ -- Energy sweep function $\overline{W}(\tau)$ of Eq.\
(\protect\ref{WbarsqrtnRes}) with $n=1$ [or of Eq.\
(\protect\ref{WbarsqrtRes})] for different values of the sweep
rate $a$.
$b$ -- The same with $n=5$.
$c$ -- The population of the exact excited level, $|c_+|^2$, for
$n=1$ and different values of $a$.
$d$ -- The same for $n=5$.
Dashed lines in all graphs correspond to $a=a_{\rm max}$ and
$\overline{W}(\tau)$ without jumps [see, e.g., Eq.\
(\protect\ref{Wtaun1a1})].
 }
\end{figure}

One of possible solutions is of the form
\begin{equation}
x=(-1)^{n}\frac{u^{n}}{\sqrt{1+u^{2n}}},\qquad u\equiv a\tau .  \label{xnSol}
\end{equation}
For $n$ odd it describes switching of the probability from $\psi
_{1}$ to $\psi _{2}$, whereas for $n$ even the system returns into
the initial state.
For $n=1$ the form of Eq.\ (\ref{xnSol}) corresponds to Eq.\
(\ref{ProbSlow}) by taking into account Eq.\ (\ref{xDef}) with
$m=|c_{1}|^{2}$.
However, the reader should note that Eq.\ (\ref{ProbSlow}) is
realized with the very slow sweep $\overline{W}(\tau )$ of an
arbitrary functional form, whereas Eq.\ (\ref{xnSol}) is valid for
a particular form of $\overline{W}(\tau )$ which is not
necessarily slow.
One obtains from Eqs.\ (\ref{yDef}) and (\ref{xnSol})
\begin{eqnarray}
\partial _{\tau }y &=&\frac{1}{\sqrt{1-x^{2}}}\partial _{\tau }x=(-1)^{n}%
\frac{anu^{n-1}}{1+u^{2n}}  \nonumber \\
\partial _{\tau }^{2}y &=&(-1)^{n}\frac{a^{2}nu^{n-2}\left[ n-1-(n+1)u^{2n}%
\right] }{\left( 1+u^{2n}\right) ^{2}}, \label{yders}
\end{eqnarray}
whereas $\tan y =x/\sqrt{1-x^{2}}=(-1)^n u^{n}$. The derivative
$\partial_{\tau }y$ has a chance to attain the value 1 at its
maximum that is determined by $\partial _{\tau }^{2}y=0$, i.e., at
\begin{equation}
u=u_{\max }=\pm \left( \frac{n-1}{n+1}\right) ^{1/(2n)}.  \label{umax}
\end{equation}
At this point one has
\begin{equation}
|(\partial _{\tau }y)_{\max }|=\left\{
\begin{array}{ll}
a, & n=1 \\
a\frac{n+1}{2}\left( \frac{n-1}{n+1}\right)^{(n-1)/(2n)}, & n>1,
\end{array}
\right.
\end{equation}
thus the condition
\begin{equation}
a\leq a_{\max }=\left\{
\begin{array}{ll}
1, & n=1 \\
\frac{2}{n+1}\left( \frac{n-1}{n+1}\right) ^{2n/(n-1)}, & n>1
\end{array}
\right.
\end{equation}
should be fulfilled. For $a<a_{\max }$ Eq.\ (\ref{WbarRes1}) yields
\begin{equation}
\overline{W}(\tau )=\frac{(-1)^{n-1}u^{n}}{\sqrt{1-\left[ anu^{n-1}/\left(
1+u^{2n}\right) \right] ^{2}}}\left[ 1-\frac{a^{2}n\left[
(2n+1)u^{2n-2}-(n-1)u^{-2}\right] }{\left( 1+u^{2n}\right) ^{2}}\right]
,\qquad u=a\tau .  \label{WbarsqrtnRes}
\end{equation}
This function is plotted for $n=1$ and $n=5$ and different values
of the parameter $a$ in Figs.\ \ref{lzi-n1and5}$a$ and
\ref{lzi-n1and5}$b$.
For $a\rightarrow a_{\max }$ discontinuities in Eq.\
(\ref{WbarsqrtnRes}) are formed at $u=\pm u_{\max }$.
These discontinuities result from the invariably negative choice
for $\cos \varphi $.
For $a=a_{\max }$ the sign of $\cos \varphi $ can change at
$u=u_{\max }$ and then these discontinuities
disappear.
This form of $\overline{W}(\tau )$ includes the factor ${\rm
sgn}\left[ \left( 1+u^{2n}\right)^{2}-\left( anu^{n-1}\right)
^{2}\right] $, in addition to Eq.\ (\ref{WbarsqrtnRes}).
Note that Eq.\ (\ref{WbarsqrtnRes}) cannot be extended into the
region $a>a_{\max }$ preserving $\overline{W}(\tau )$ real, in
contrast to Eq.\ (\ref{xnSol}).
The small- and large-argument forms of $\overline{W}(\tau )$ are
\begin{equation}
\overline{W}(\tau )\propto (-1)^{n-1}\left\{
\begin{array}{ll}
u^{n}, & |u|\gg 1 \\
u^{n-2}, & |u|\ll 1,\qquad n>1
\end{array}
\right. .  \label{WnLims}
\end{equation}
In the case $n=1$, Eq.\ (\ref{WbarsqrtnRes}) simplifies to
\begin{equation}
\overline{W}(\tau )=\frac{u}{1+u^{2}}\frac{1-3a^{2}+2u^{2}+u^{4}}{\sqrt{%
1-a^{2}+2u^{2}+u^{4}}}.  \label{WbarsqrtRes}
\end{equation}
One can see that for $u\gg 1$ the function $\overline{W}(\tau )$ describes a
linear energy sweep with the velocity $v=a$.
For $u\sim 1$ the form of $\overline{W}(\tau )$ is nonlinear and
for $a<a_{\max }=1$ the resonance is crossed three times,
including $u=0$.
For $a=1$ Eq.\ (\ref{WbarsqrtRes}) yields the even function
\begin{equation}
\overline{W}(\tau )=\frac{2-2\tau ^{2}-\tau ^{4}}{(1+\tau ^{2})\sqrt{2+\tau
^{2}}}  \label{Wtaun1a1}
\end{equation}
This sweep crosses the resonance two times, in the forward and
backward directions, which means that the system is in the {\em
excited} state $\psi_2$ at the end of the sweep at $t=\infty$.
For $a<1$ and far from the resonance Eq.\ (\ref{WbarsqrtRes})\ \
has the form
\begin{equation}
W(t)=\Delta \overline{W}(\tau )=\Delta a\tau =\left( \Delta
^{2}a/\hbar \right) t,
\end{equation}
thus the corresponding linear sweep rate is $v=\Delta^{2}a/\hbar$.
The maximal sweep rate that ensures the complete switching to
$\psi _{2}$ is realized at $a=1$ is $v_{\max }=\Delta ^{2}/\hbar$.
For a comparison, in the regular LZ effect with the probability to
stay in the initial state given by Eq.\ (\ref{LZProbab}) the sweep
rate $v\ll \pi \Delta ^{2}/(2\hbar ) $ is required for a complete
switching.
For $v=\Delta^{2}/\hbar $ the probability to stay in the initial
state is still $P=0.208$.

Another solution that describes switching from state 1 to state 2
is
\begin{equation}
x=-\tanh u,\qquad u=a\tau .
\end{equation}
Here the sweep function determined from Eq.\ (\ref{WbarRes1}) reads
\begin{equation}
\overline{W}(\tau )=\frac{\sinh u}{\sqrt{1-a^{2}/\cosh ^{2}u}}\left[ 1-\frac{%
2a^{2}}{\cosh ^{2}u}\right] .  \label{Wtaua}
\end{equation}
In the range $1/\sqrt{2}<a<1$ the function $\overline{W}(\tau )$
changes its sign three times and its derivative at $\tau =0$ is
negative.
With $a\to 1$ this derivative increases unlimitedly and a jump of $%
\overline{W}(\tau )$ at $\tau =0$ is formed.
For $a=a_{\max }=1$ the solution with eliminated jump is an even
function
\begin{equation}
\overline{W}(\tau )=\frac{2}{\cosh \tau }-\cosh \tau .  \label{Wtau}
\end{equation}
This sweep function crosses the resonance two times.
 Again, Eq.\ (\ref {Wtaua}) cannot be continued into the region $a>1$.

Now we return to the solution for the probability $x$ described by Eq.\ (\ref
{xnSol}) and analyze its behavior for large $n$.
It has a form of three plateaus $x\cong 1$ for $u\lesssim -1$,
$x\cong 0$ for $-1\lesssim u\lesssim 1$, and $x\cong -1$ for
$1\lesssim u$ with two narrow transition regions between them.
In the central region $-1\lesssim u\lesssim 1$ one has
$|c_{2}|^{2}\cong |c_{1}|^{2}=(1+x)/2\cong 1/2$, i.e., the system
is in the mixture of the states $\psi _{1}$ and $\psi _{2}$ with
equal weights.
The sweep function $\overline{W}(\tau )$ of Eq.\
(\ref{WbarsqrtnRes}) is fast approaching the resonance at $u\cong
-1$ and remaining in the close
vicinity of the latter until $u\cong 1$.
Since the probabilities $|c_{1}|^{2}\cong |c_{2}|^{2}\cong 1/2$ do
not oscillate in the range $-1\lesssim u\lesssim 1$ (unlike the
case in which the system is instantaneously brought into
resonance), one is lead to the conclusion that the system is in
one of the exact quantum states $\psi _{-}$ or $\psi _{+}$.
It is naturally of interest to question in which of these exact
states does the system come under the influence of the energy
sweep described by Eq.\ (\ref{WbarsqrtnRes}).
We will see below that both variants can be realized.
This opens a possibility to prepare the system in the ground or
excited state in
resonance by the appropriate choice of $\overline{W}(\tau )$.
Indeed, for $n\geq 3$ one has $\overline{W}(\tau )=\partial_{\tau
}y=\partial _{\tau }^{2}y=0$ at $\tau =0, $ thus from Eq.\
(\ref{ypsiEqRed})\ one can see that the sweep can be stopped at
$\tau =0$ and the system will remain in this state.
To demonstrate this behavior one should rewrite the wave function
of the system $\psi =c_{1}(t)\psi_{1}+c_{2}(t)\psi_{2}$ determined
above in the basis of the exact states
\begin{equation}
\psi_{\pm }(t)=\frac{1}{\sqrt{2}}\left[ \pm K_{\pm }(t)\psi
_{1}+K_{\mp }(t)\psi _{2} \right] ,\qquad K_{\pm }(t)\equiv
\sqrt{1\pm \frac{\overline{W}(t)}{\sqrt{1+ \overline{W}^{2}(t)}}}
\label{psipm}
\end{equation}
corresponding to the time-dependent ``energy eigenvalues'' of Eq.\
(\ref{epspmDef})\ as $\psi =c_{+}(t)\psi _{+}(t)+c_{-}(t)\psi
_{-}(t)$.
The corresponding probabillities are given by
\begin{equation}
|c_{\pm }(t)|^{2}=\frac{1}{2}\left\{ 1\pm
\frac{\overline{W}(t)x(t)+\sqrt{1-x^{2}(t)} \cos \varphi(t)
}{\sqrt{1+\overline{W}^{2}(t)}}\right\} .  \label{cpm2}
\end{equation}
One can see that in resonance at $t=0$ ($\overline{W}=x=0$, $\cos
\varphi =\pm 1$) the situation depends on the sign of $\cos
\varphi$.
For sweep functions given by Eq.\ (\ref{WbarsqrtnRes}) with
$a<a_{\max }$ one has $\cos \varphi =-1$ and the system is in the
ground state ($|c_{-}(t)|^{2}=1$, $ |c_{+}(t)|^{2}=0)$.
For the special sweep functions with $a=a_{\max }$ one has $\cos
\varphi =1$ and the system is in the excited state
($|c_{-}(t)|^{2}=0,$ $|c_{+}(t)|^{2}=1)$, the transition from the
ground state $\psi_{-}(t )$ to the excited state $\psi _{+}(t) $
occuring for $n\gg 1$ in the vicinity of $u=u_{\max}$ of Eq.\
(\ref{umax}).
These features are illustrated for the sweep functions of Eqs.\
(\ref{WbarsqrtnRes}) and (\ref{Wtaun1a1}) in Figs.\
\ref{lzi-n1and5}$c$ and \ref{lzi-n1and5}$d$.
Note that for $n=1$ and $a=a_{\max}=1$ the sweep function of Eq.\
(\ref{Wtaun1a1}) returns to $-\infty$ at $\tau=\infty$ and the
system ends up in the excited state $\psi_+=\psi_2$.

To conclude, we have solved the inverse Landau-Zener problem for a
two-level system, i.e., we have established the form of the sweep
function $W(t)\cong \varepsilon_1(t)-\varepsilon_2(t)$ needed to
insure a given probability $|c_1(t)|^2$ to be in the state 1 at
time $t$.
We have discussed two examples for which the system performs a
{\em complete} transition from 1 to 2.
In contrast to the direct LZ problem with a linear sweep, this can
be accomplished by a {\em finite} sweeping rate.
Further we have found such sweep functions that prepare the system
in its exact ground state or in its exact excited state at
resonance.
It is not difficult to find $W(t)$ that prepare the system in an
{\em arbitrary} mixture of the exact states at resonance, which
can be potentially used in quantum computing based on tunneling
units.
%


\begin{thebibliography}{1}

\bibitem{lan32}
{L. D. Landau}, Phys. Z. Sowjetunion {\bf 2},  46  (1932).

\bibitem{zen32}
C. Zener, Proc. R. Soc. London A {\bf 137},  696  (1932).

\bibitem{dobzve97}
{V. V. Dobrovitski and A. K. Zvezdin}, Europhys. Lett. {\bf 38},
377  (1997).

\bibitem{lanlif3}
{L. D. Landau and E. M. Lifshitz}, {\em Quantum {M}echanics}
(Pergamon, London,
  1965).

\bibitem{crohug77}
{D. S. F. Crothers and J. G Huges}, J. Phys. B {\bf 10},  L557
(1977).

\bibitem{werses99}
{W. Wernsdorfer and R. Sessoli}, Science {\bf 284},  133  (1999).

\bibitem{losdivigri92}
{D. Loss, D. P. DiVincenzo, and G. Grinstein}, Phys. Rev. Lett.
{\bf 69},  3232
   (1992).

\bibitem{delhen92}
{J. von Delft and C. L. Henley}, Phys. Rev. Lett. {\bf 69},  3236
(1992).

\bibitem{garg93}
{A. Garg}, Europhys. Lett. {\bf 22},  205  (1993).

\end{thebibliography}

\end{document}